\documentclass[final]{ias2}
\usepackage{graphicx} 
\usepackage{multirow}
\usepackage{array} 

\usepackage{hyperref} 
\def\stat{\mbox{$\;$(stat.)}}
\def\syst{\mbox{$\;$(syst.)}}
\def\statsyst{\mbox{$\;$(stat.+syst.)}}
\def\lumi{\mbox{$\;$(lumi.)}}

\def\met{\ensuremath{E_{\mathrm{T}}^{\mathrm{miss}}}}
\def\ttbar{\ensuremath{t\bar{t}}}
\def\stt{\ensuremath{\sigma_{t\bar{t}}}}
\def\st{\ensuremath{\sigma_{t}}}
\def\mtop{\ensuremath{m_{\mathrm{top}}}}
\def\pt{\ensuremath{p_{\mathrm{T}}}}
\def\TeV{\ifmmode {\mathrm{\ Te\kern -0.1em V}}\else
                   \textrm{Te\kern -0.1em V}\fi}%
\def\GeV{\ifmmode {\mathrm{\ Ge\kern -0.1em V}}\else
									\textrm{Ge\kern -0.1em V}\fi}%
\def\ipb{\mbox{pb$^{-1}$}}%  Inverse picobarns.
\def\ifb{\mbox{fb$^{-1}$}}%  Inverse picobarns.
\def\bfig{\begin{figure*}[htbp]\centering}
\def\efig#1#2{\caption{#2}\label{fig:#1}\end{figure*}}
\def\igra#1#2{\raisebox{-0.5\height}{\includegraphics[width=#2\textwidth]{#1}}}

\def\threeplot#1#2#3#4#5#6#7#8{\bfig\igra{#3}{#4}\igra{#5}{#6}\igra{#7}{#8}\efig{#2}{#1}}
\def\conf#1#2{\bibitem{c#1}The ATLAS Collaboration, ATLAS-CONF-2011-#1, cdsweb.cern.ch/record/#2}

\begin{document}

\markboth{Top physics at ATLAS}{Markus Cristinziani}

\title{Top physics with 0.70--1.08 $\ifb$ of $\mathbf{pp}$ collisions with the ATLAS detector at the LHC}

\author[*]{Markus Cristinziani\footnote{on behalf of the ATLAS Collaboration}}
\email{cristinz@uni-bonn.de}
%\address[pi]{on behalf of the ATLAS Collaboration}
%Physikalisches Institut, Universit\"{a}t Bonn, Nussallee 12, 53127 Bonn, Germany}

\begin{abstract}
With data collected during the first half of the 2011 $pp$ run of the Large Hadron Collider at $\sqrt{s} = 7 \TeV$,
a substantial data sample of high $\pt$ triggers, corresponding to an integrated luminosity of $1.08 \ifb$, 
has been collected by the ATLAS detector. 
Measurements of the production of top-quark pairs and single top-quarks in different channels,
the top-quark mass, the top-quark pair charge asymmetry and spin correlations, and the $W$ helicity fractions in
top-quark decays are presented, as well as two searches for new physics effects involving top-quark pairs.

{\em Proceedings of the Lepton Photon 2011 Conference, to appear in Pramana - journal of physics}
\end{abstract}

\keywords{Top quarks, QCD experimental tests, Kaluza-Klein excitations}

\pacs{14.65.Ha, 12.38.Qk, 14.80.Rt, 14.80.Ly}
 
\maketitle

\section{Introduction}

Top-quark measurements are of central importance to the LHC physics programme.
The production of top-quark pairs in $pp$ collisions is a process which is situated 
at the boundary between the Standard Model (SM) and what might lie beyond it.
Within the SM, top quarks are predicted to almost always decay to a $b$-quark and a 
$W$-boson, that can further decay leptonically or hadronically. The final
states are thus characterised by the presence of two (dilepton), one (single lepton)
or no leptons (all hadronic channel). 
With an integrated luminosity of $0.70 - 1.08 \ifb$ of $pp$ collisions recorded by
the ATLAS detector at $\sqrt{s} = 7 \TeV$ in the first half of 2011 a variety of 
precision and new measurement have been obtained.

%$, in the realm of the top-quark pair production 
%cross-section, the search and measurement of single-top production, properties, such
%as the top quark mass, the top-quark pair charge asymmetry and spin correlation,
%the W helicity and searches for new physics signatures with additional \met or \ttbar
%resonances.

\section{Top quark pair production cross-section $\stt$}
Within the SM the $\ttbar$ production cross-section at $\sqrt{s} = 7 \TeV$
is calculated to be \mbox{$165^{+11}_{-16}$ pb} at approximate NNLO for 
a top-quark mass of $172.5 \GeV$. 
A precise determination of $\stt$ tests these perturbative QCD predictions.
First measurements of $\stt$ at the LHC have been reported by ATLAS and CMS
with 3 $\ipb$ and 35 $\ipb$.
With approximately twenty times more data the single lepton, dilepton and lepton+$\tau$
channels have been measured.
In the single-lepton channel a multivariate likelihood discriminant is constructed using
template distributions of four variables~\cite{c121}. Data are split according to lepton 
flavour and number of jets.  A profile likelihood
technique is used to extract $\stt$ and constrain the systematic effects from data.
The result is $\stt = 179.0 \pm 9.8 \statsyst \pm 6.6 \lumi$ pb, the most
precise measurement of $\stt$ at LHC to date.
In the dilepton channel a cut-and-count
method is employed~\cite{c100}. 
The background contribution from 
Drell-Yan production is suppressed by requiring 
for same-flavour events large $\met$ and 
for $e\mu$ events large $H_T$, 
the scalar sum of jet and lepton transverse energies.
Remaining Drell-Yan events and
background from fake leptons are estimated with data-driven methods.
The combined cross-section in the three channels is
$\stt = 171 \pm 6 \stat ^{+16}_{-14} \syst \pm 8 \lumi$ pb.
A second measurement requiring at least one $b$-jet yields a consistent result.
A cross-section measurement is also performed in final states 
with an isolated muon and a hadronically decaying $\tau$-lepton~\cite{c119}.
At least two jets, one of them $b$-tagged, are required.
The analysis utilises a multivariate 
technique based on boosted decision trees to identify 
$\tau$-leptons and extracts $\stt = 142 \pm 21 \stat^{+20}_{-16} \syst \pm 5 \lumi$ pb.
\threeplot{Data superimposed on expectations for the measurement of the top-quark 
pair production cross-section in the single-lepton~\cite{c121}, dilepton~\cite{c100} and 
lepton+$\tau$~\cite{c119} channels, respectively.}{xsplot}{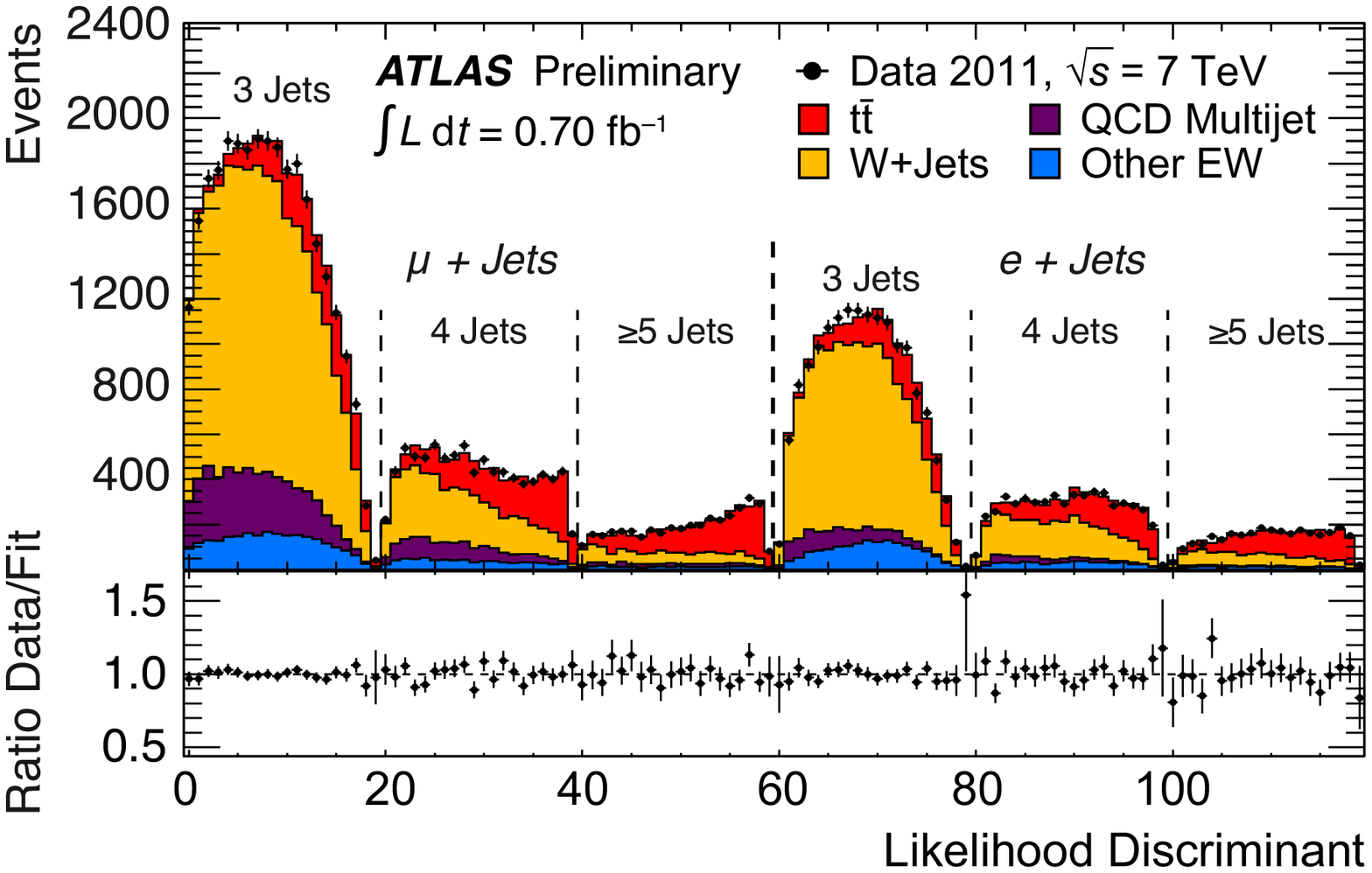}{0.48}{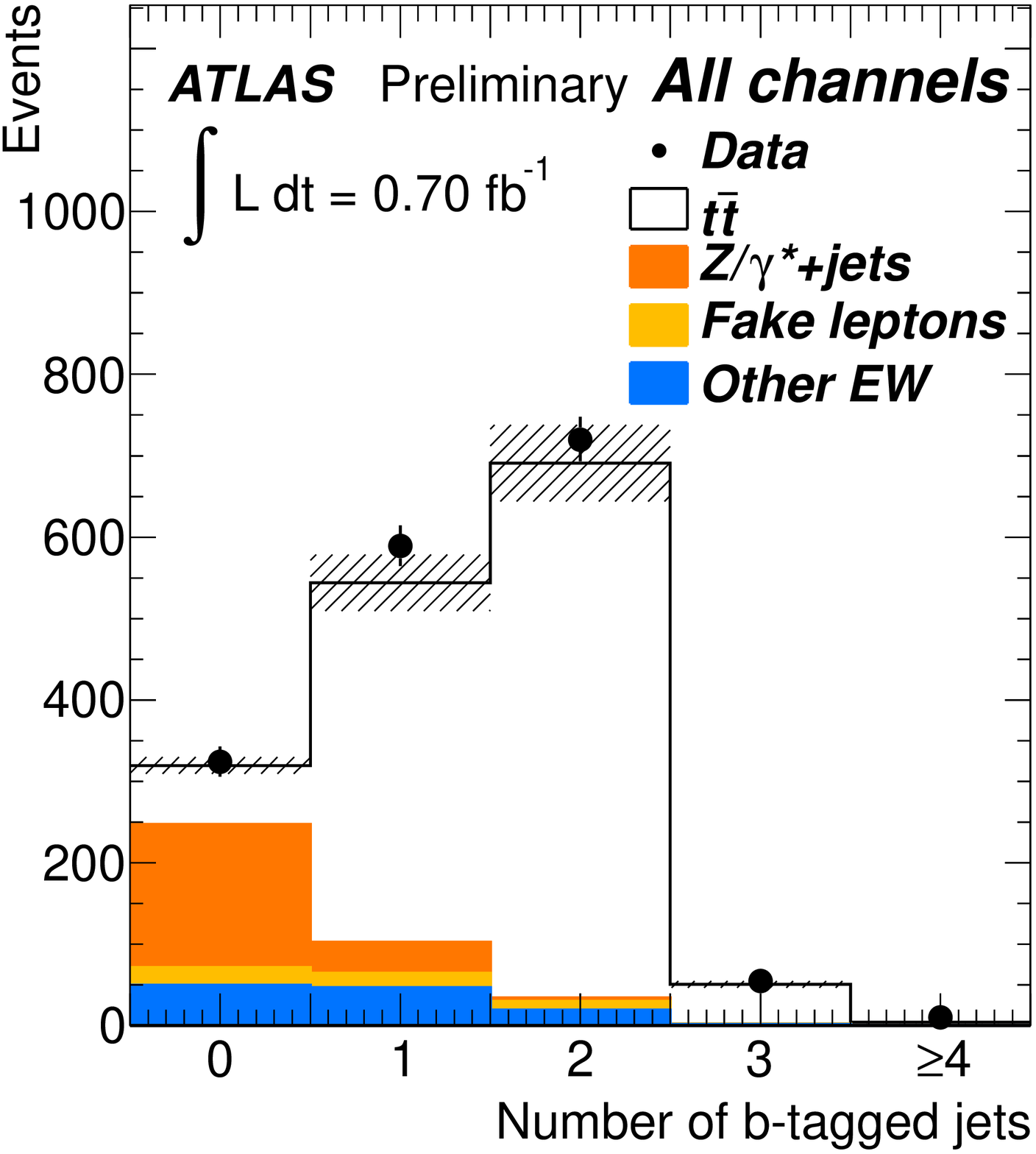}{0.26}{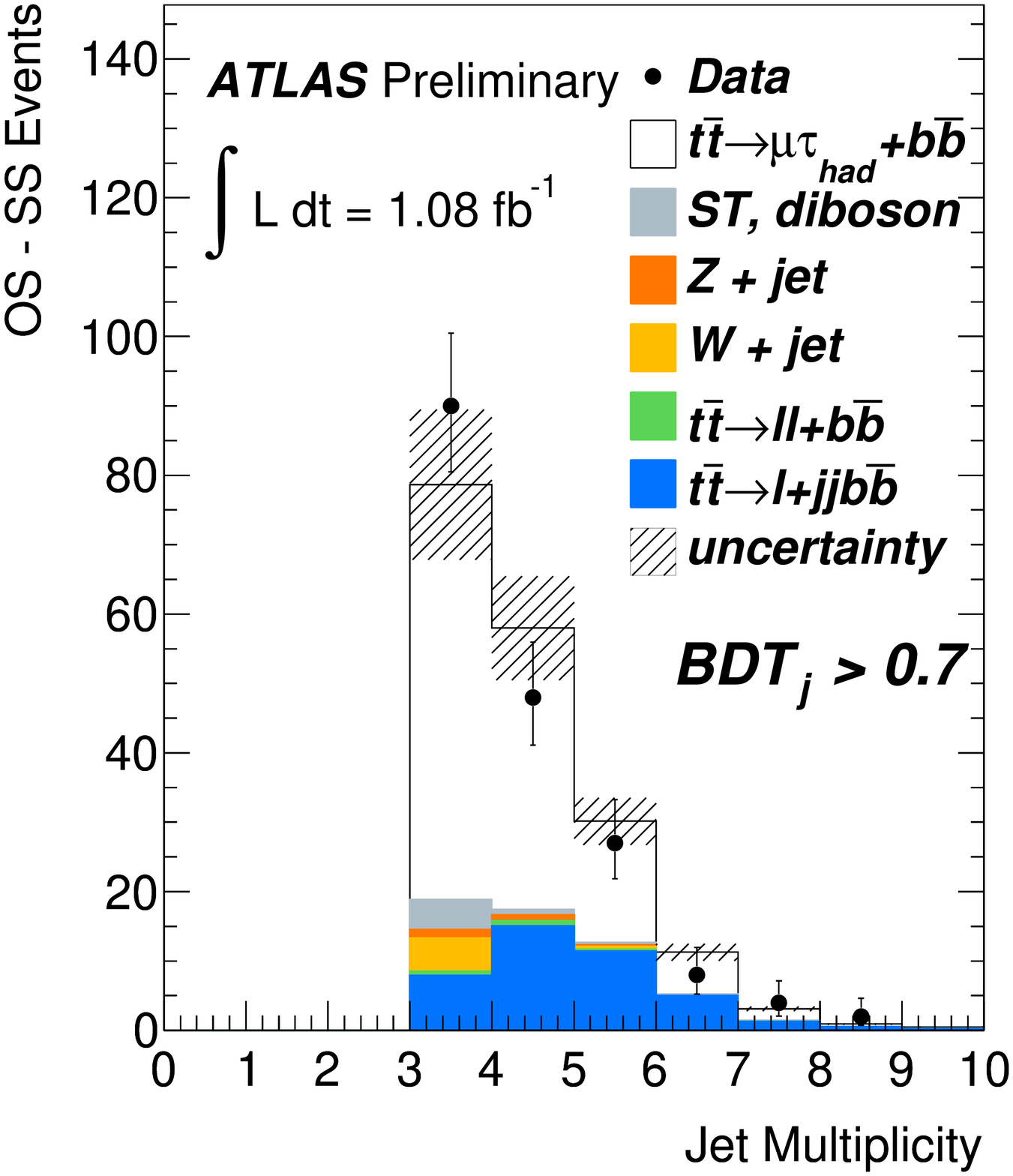}{0.25}

\section{Single top-quark production}
Single top-quark production has been explored in the three channels ($t$, $Wt$ and $s$).
For the $t$-channel measurement a cut-based analysis exploiting the two- and three-jet bins
is performed and exactly one $b$-tag is required~\cite{c101}.
An excess of top candidates is observed over antitop candidates.
The measured cross-section is $\st^\mathrm{t-chan} = 90 \pm 9 \stat^{+31}_{-20} \syst $ pb in
agreement with the SM prediction and has been cross-checked with an analysis based on 
neural networks.
For the search for the associated production of a top quark and a $W$-boson 
only the leptonic decays of the two $W$ bosons are considered~\cite{c104}.
A simple cut-based approach is used to select the $Wt$ contribution. 
At the 95\% C.L. $\st^\mathrm{Wt-chan} < 39 (41)$ pb for the observed (expected) upper limit.
For the search in the s-channel single top-quark production two b-jets are required~\cite{c118}.
With a cut-based analysis an upper limit of 
$\st^\mathrm{s-chan} < 26.5 (20.5)$ pb is obtained at 95\% C.L.
\bfig\igra{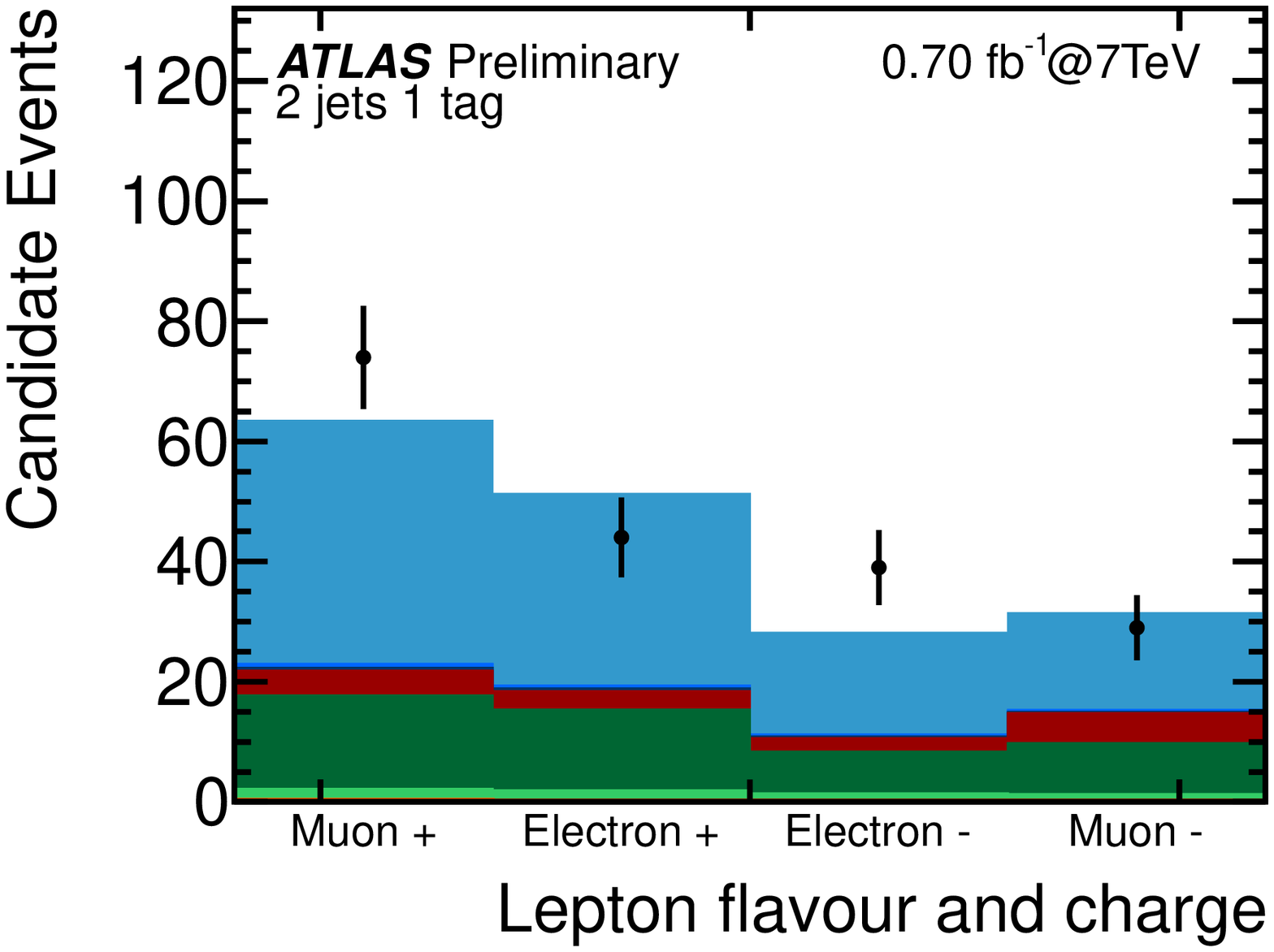}{0.32}\igra{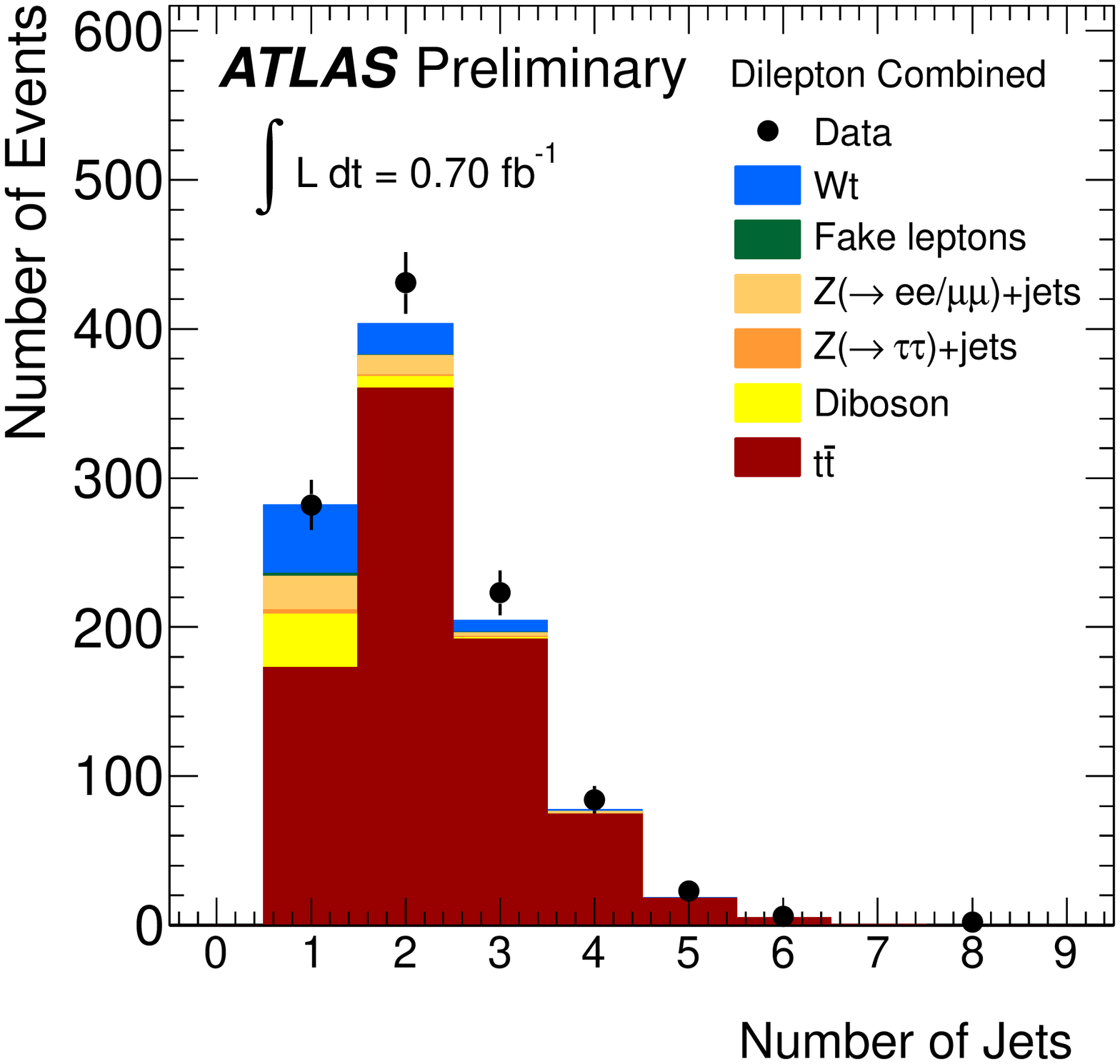}{0.25}\igra{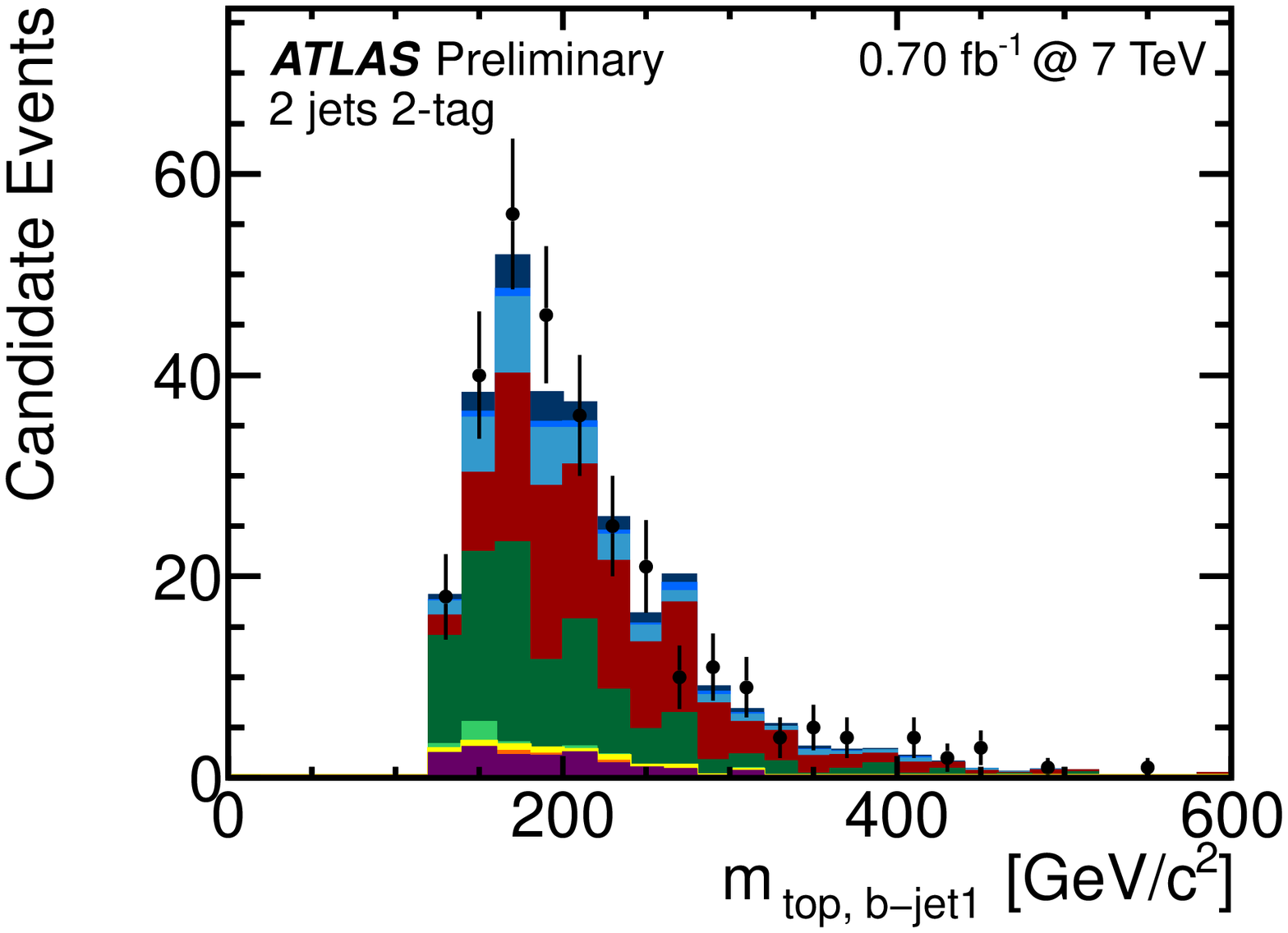}{0.32}\igra{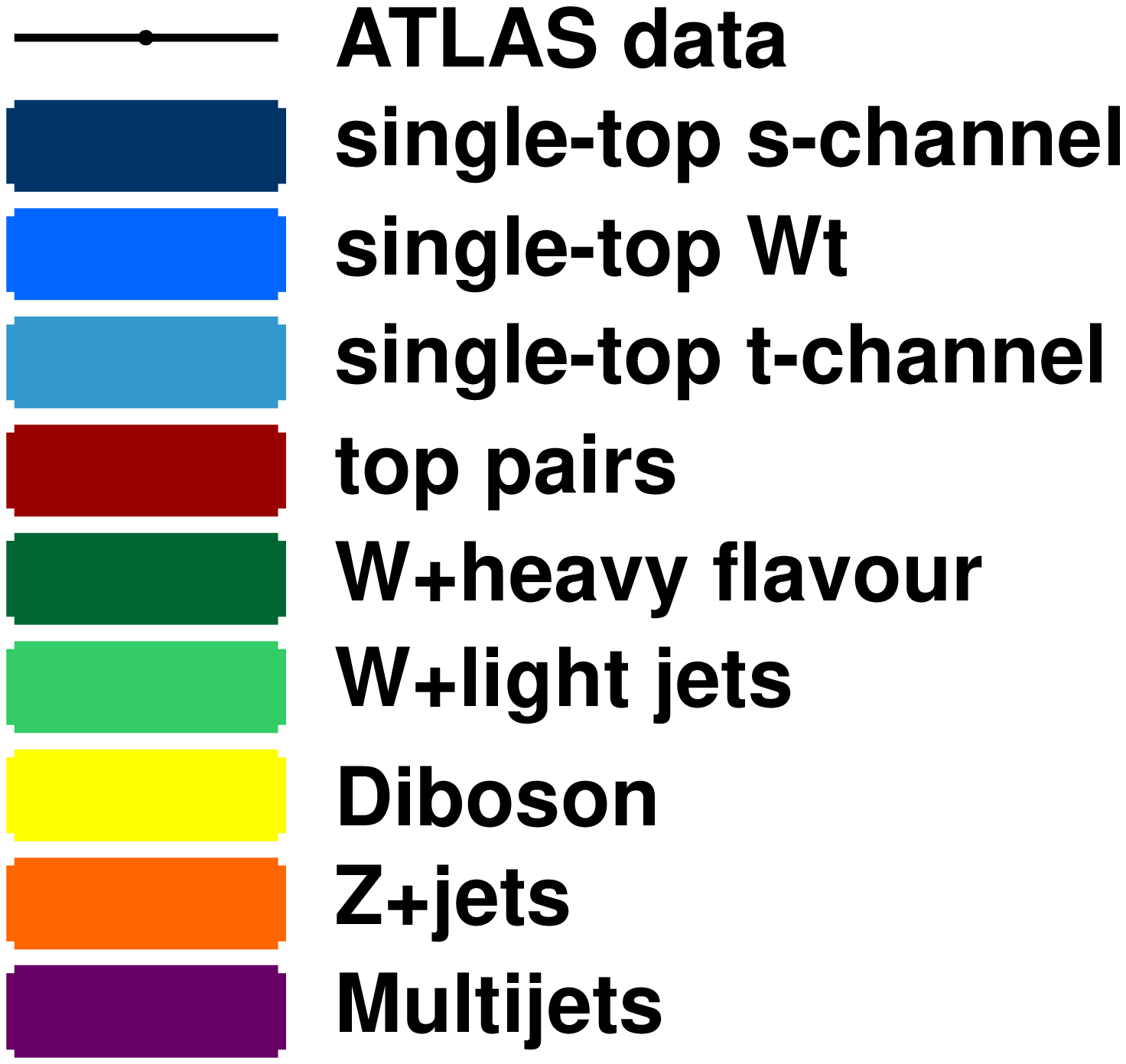}{0.15}\efig{stop}%
{Single top-quark production in the $t$-(left~\cite{c101}), $Wt$-(centre~\cite{c104}), %
and $s$-channel~(right~\cite{c118}), respectively}

\section{Top-quark properties}
The top-quark mass ($\mtop$) has been measured using 
a two-dimensional template analysis in the single-lepton channel
that determines $\mtop$ together with a global jet energy 
scale factor~\cite{c120}. 
Combining the electron and muon channels and the results from the 2010 ATLAS data, 
the top-quark mass is measured to be $\mtop = 175.9 \pm 0.9 \stat \pm 2.7 \syst \GeV$.
The top-antitop charge asymmetry is measured in the single-lepton channel~\cite{c106}.
A kinematic likelihood is used to reconstruct the \ttbar~event topology.  After background
subtraction, a Bayesian unfolding procedure is performed to correct
for acceptance and detector effects. The charge asymmetry observable
$A_{\textrm{C}}$ is based on the difference of the absolute values of top and antitop rapidities,
$|Y_t| - |Y_{\bar{t}}|$. It is measured to be
$A_{\textrm{C}}  =  -0.024 \pm 0.016 \stat \pm 0.023 \syst$,
in agreement with the SM prediction of $A_{\textrm{C}} = 0.006.$
Spin correlation in $\ttbar$ events are studied in the dilepton channel~\cite{c117}.
The spin information is accessed via the angular distributions of its decay products. 
The difference in azimuthal angle between the two charged leptons is 
compared to the expected distributions in the SM, and to the case where the 
top quarks are produced with uncorrelated spin. Using the helicity basis as the quantisation axis, 
the strength of the spin correlation between the top and antitop quark is measured 
to be $A_\mathrm{helicity}=0.34^{+0.15}_{-0.11}$, which is in agreement with the NLO SM 
prediction.
Helicities of $W$-bosons in top-quark decays have been measured in the single- and di-lepton channels~\cite{c122}.
At least one of the jets in the single-lepton final state is required to be $b$-tagged. 
The results are in agreement with the SM. As the 
polarisation of the $W$-bosons in top-quark decays is sensitive to the structure of the 
$Wtb$-vertex the measurements are used to set limits on anomalous contributions to the $Wtb$-vertex.
\threeplot{Summary of ATLAS top-quark mass measurements (left~\cite{c120}), unfolded $\Delta |Y|$ 
distribution for the $e$+jets channel in the charge asymmetry measurement (centre~\cite{c106}) 
and $\Delta \varphi$ of the two leptons for all dilepton channels in the spin correlation 
measurement (right~\cite{c117}).}{prop}{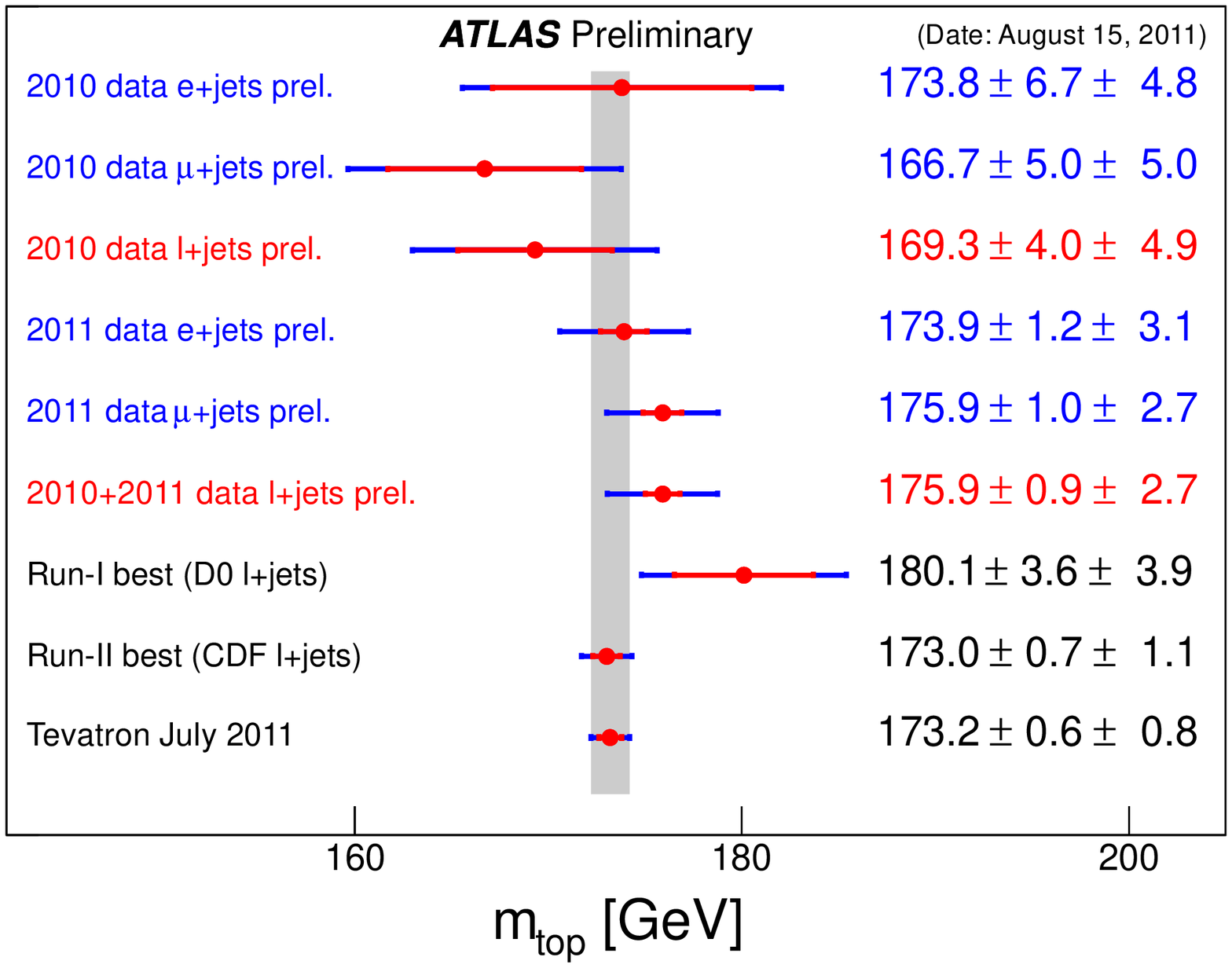}{0.39}{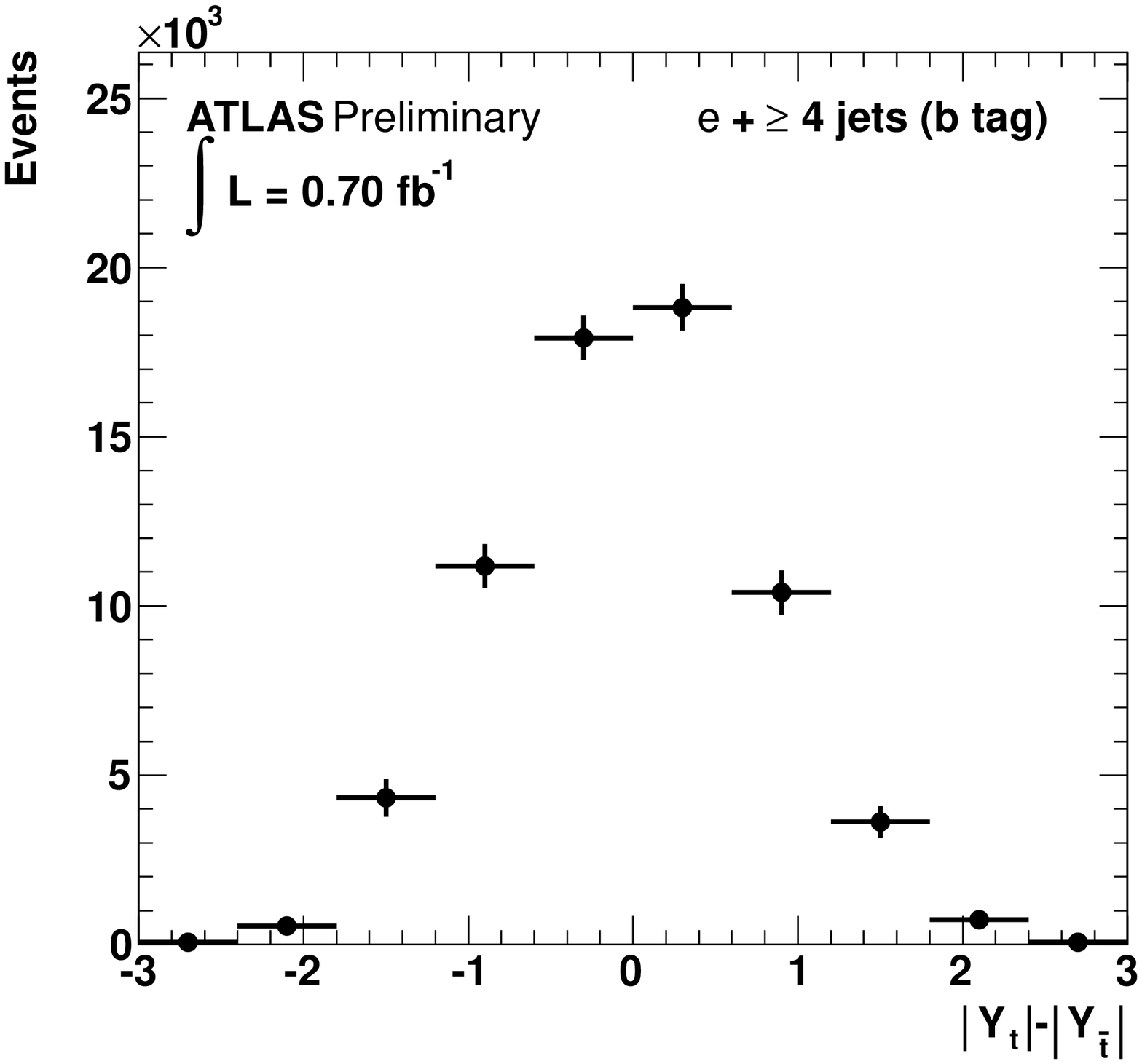}{0.34}{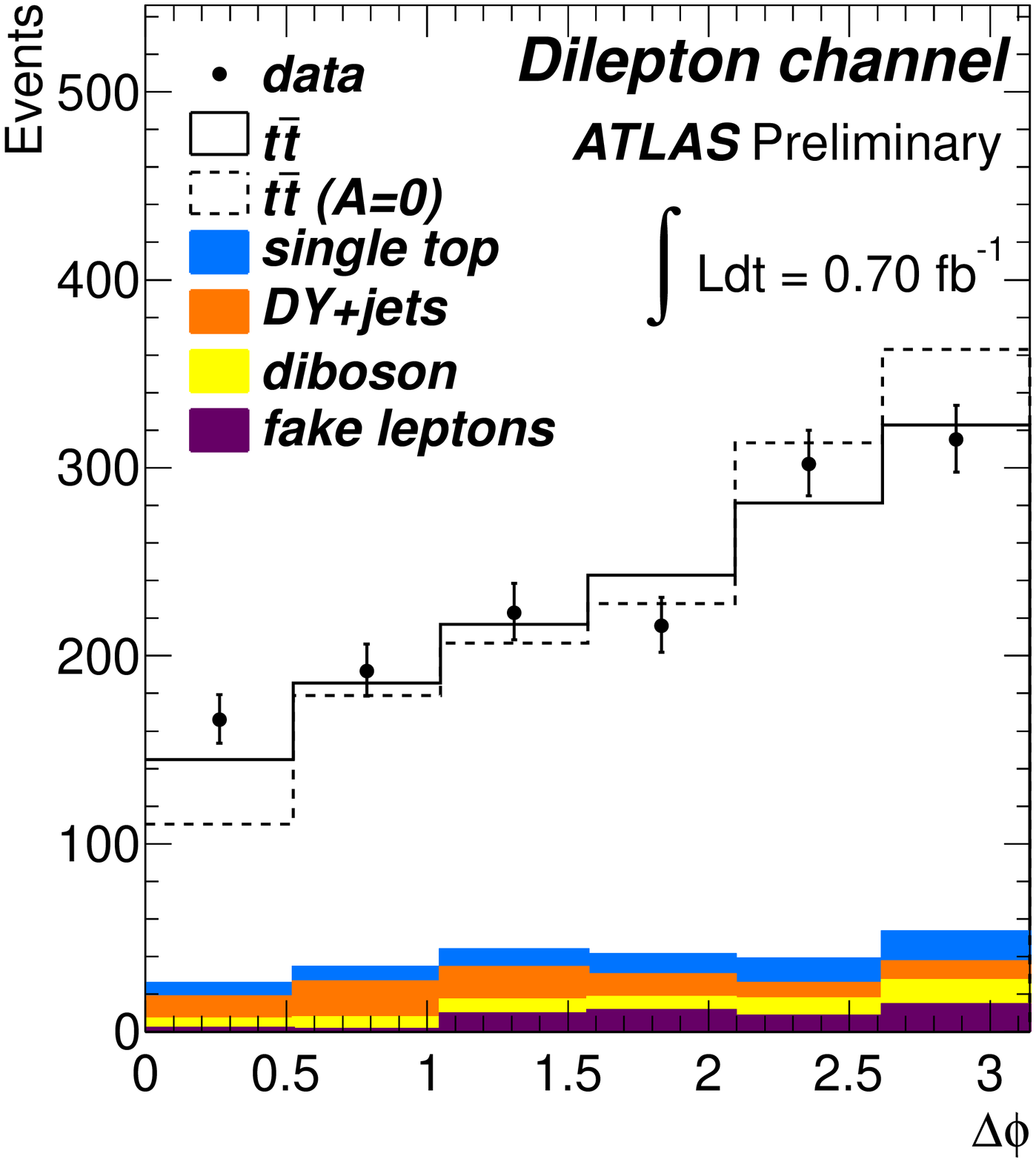}{0.27}

\section{Search for new physics in top-quark pair final states} 
A search for new phenomena in $\ttbar$ events with large $\met$ is carried out in 
the single-lepton channel~\cite{etmiss}. 
The results are interpreted in terms of a model where new top-quark 
partners are pair-produced and each decay to a top quark and a long-lived 
undetected neutral particle. The data are found to be consistent with SM expectations. 
A limit at 95\% C.L. is set excluding a cross-section times branching ratio of 
1.1 pb for a top-partner mass of 420 $\GeV$ and a neutral particle mass less than 10 $\GeV$. 
In a search for high mass $\ttbar$ resonances in the dilepton channel~\cite{c123} no
excess above the SM expectation is observed. 
Upper limits at the 95\% C.L are set on the cross-section times 
branching ratio of the resonance decaying to $\ttbar$ pairs as a function of the resonance 
pole mass. A lower mass limit of 0.84 $\TeV$ is set for the case of a Kaluza-Klein 
gluon resonance in the Randall-Sundrum Model.
\threeplot{Limits on anomalous couplings in the $Wtb$ vertex (left~\cite{c122}), on pair-produced vector
'quark' partner T of the top quark (centre~\cite{etmiss}) and Kaluza-Klein gluons from a search for 
resonances in the $m_{\ttbar}$ spectrum in the dileptonic final state (right~\cite{c123}).}
{search}{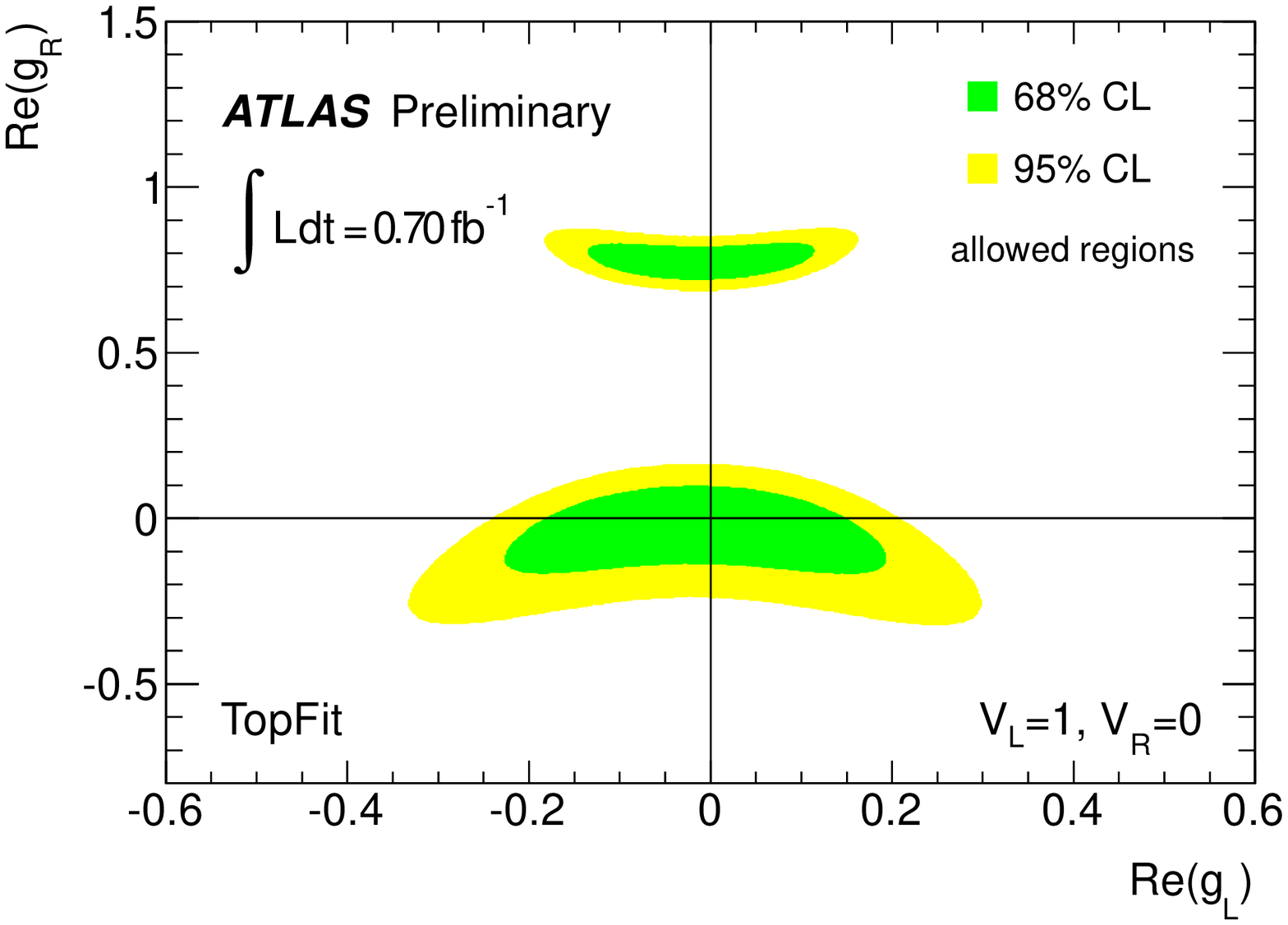}{0.33}{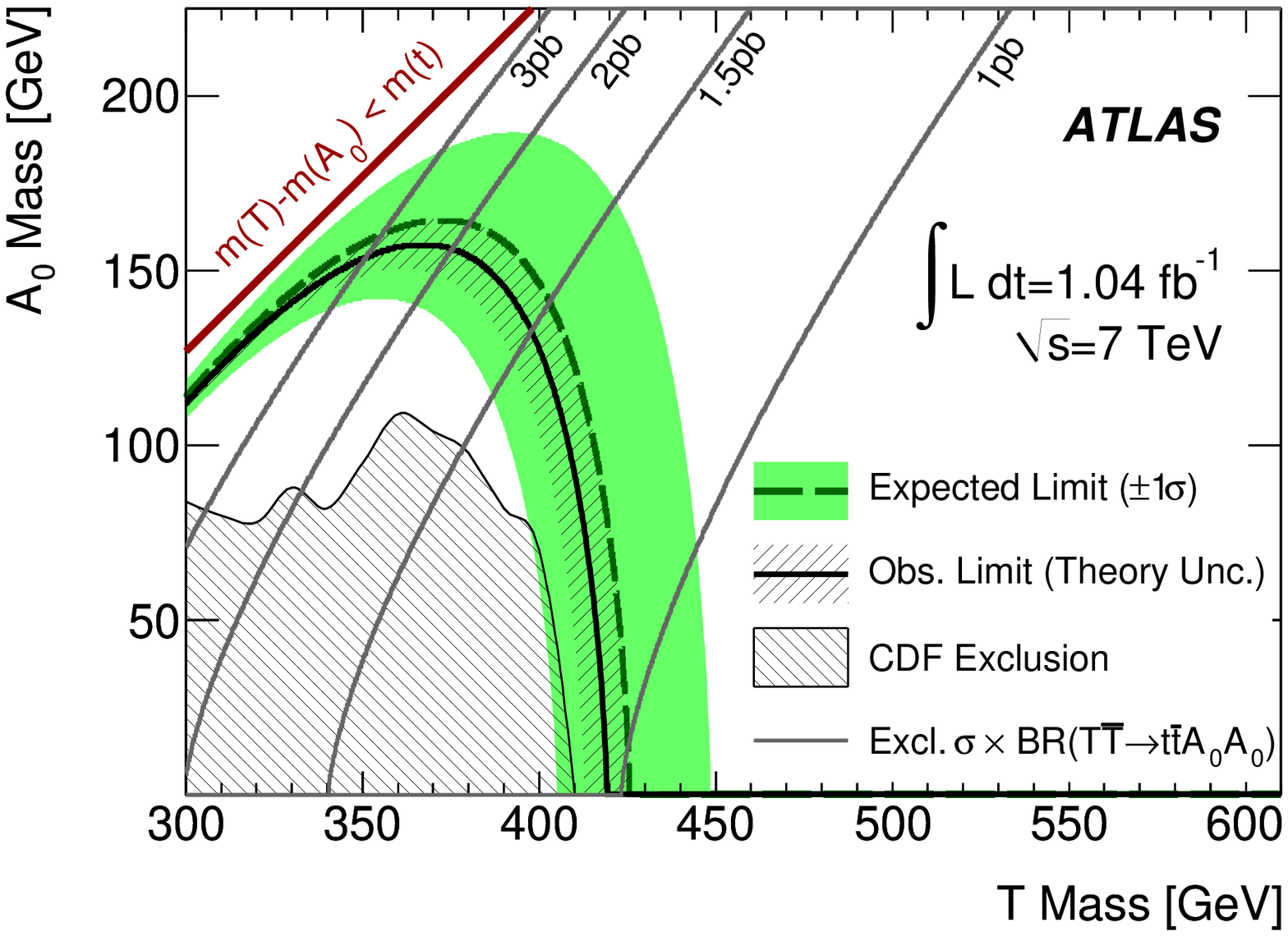}{0.33}{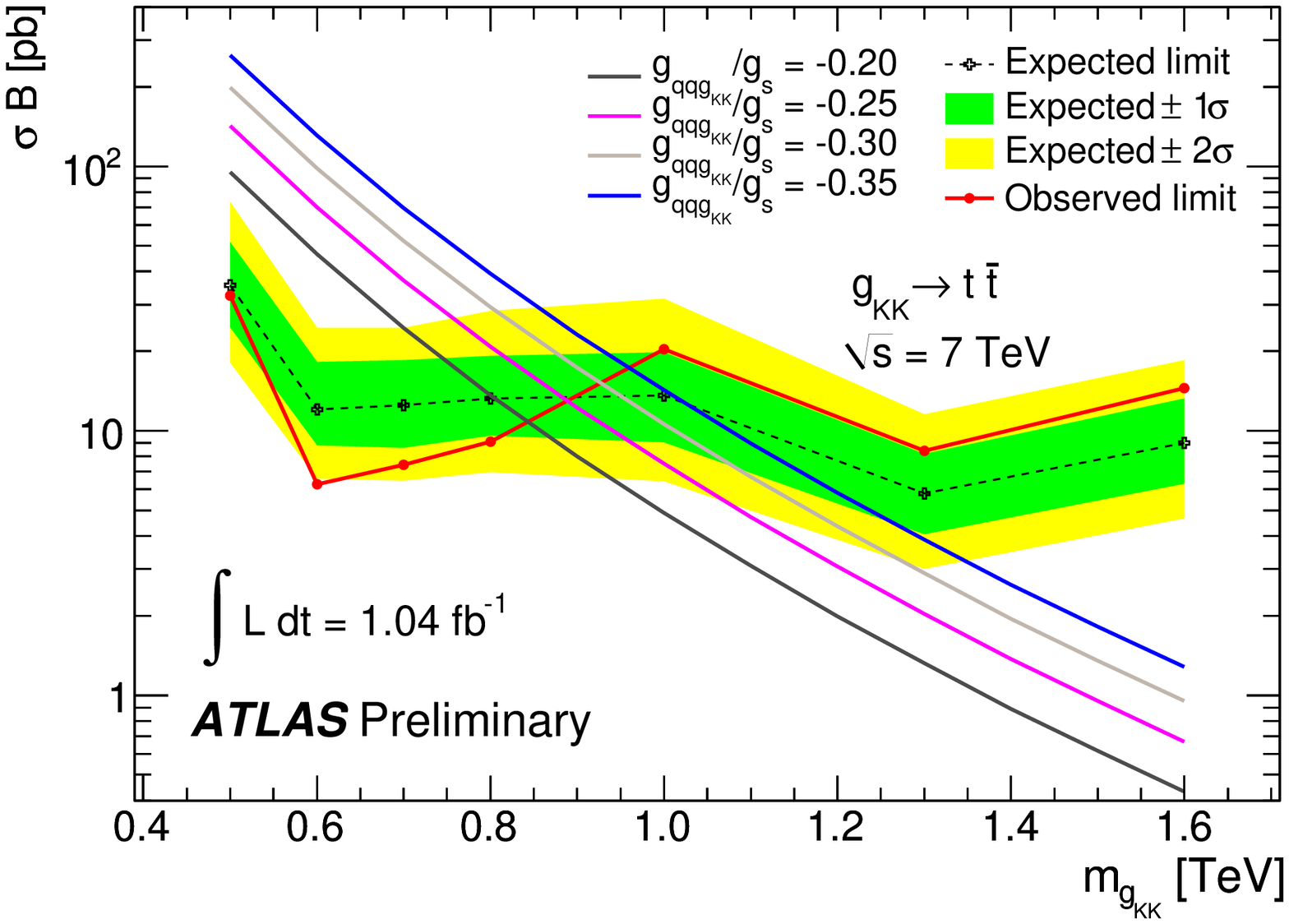}{0.33}
\acknowledgments
I would like to thank the LHC crew and the ATLAS Collaboration for the excellent data quality,
the many authors and contributors of the top physics analyses shown here, 
David Milstead for 
reading the manuscript and the organizers of Lepton Photon 2011 in Mumbai for the excellent organisation of 
the conference.
I also gratefully acknowledge the support of the Deutsche 
Forschungsgemeinschaft (DFG) through the Emmy-Noether grant CR-312/1-1.

\end{document}

